\documentclass[12pt]{article}
\usepackage{amsmath}
\usepackage{amssymb}
\usepackage{cancel}
\usepackage{graphicx}
\usepackage{color}
\usepackage{wrapfig}

\renewcommand{\thefootnote}{\fnsymbol{footnote}}

\begin{document}
\baselineskip=19.5pt

\begin{titlepage}

\begin{center}
\vspace*{17mm}

{\large\bf%
Dark matter and $U(1)'$ symmetry for the right-handed neutrinos
}

\vspace*{10mm}
Manfred~Lindner,
\footnote[1]{~lindner@mpi-hd.mpg.de}
~~Daniel~Schmidt,
\footnote[2]{~daniel.schmidt@mpi-hd.mpg.de}
~~Atsushi~Watanabe
\footnote[3]{~atsushi.watanabe@mpi-hd.mpg.de}
\vspace*{10mm}

{\small {\it 
Max-Planck-Institut f\"ur Kernphysik, Saupfercheckweg 1, 69117 Heidelberg, Germany
}}\\

\vspace*{3mm}

{\small (October, 2013)}
\end{center}

\vspace*{7mm}

\begin{abstract}\noindent%
We consider a $U(1)'$ gauge symmetry acting on three generations of right-handed 
neutrinos.
The $U(1)'$ symmetry is broken at the TeV scale and its remnant discrete symmetry
makes one of the right-handed neutrinos stable.
As a natural consequence of the anomaly cancellation,
the neutrino mass matrix consists of a combination of Type I (TeV scale) seesaw and 
radiative corrections. 
The stable right-handed neutrino communicates with the Standard Model
via $s$-channel exchange of the Higgs field and the $U(1)'$ gauge boson, so that 
the observed relic density for dark matter is obtained in a wide range of the parameter 
space.
The experimental signatures in collider and other experiments are briefly discussed.
\end{abstract}

\end{titlepage}

\newpage
\renewcommand{\thefootnote}{\fnsymbol{footnote}}
\section{Introduction}
\label{intro}
Do the physics of neutrino mass and dark matter share the same origin ?
The experimental evidence for neutrino mass~\cite{SK,SNO,KamLAND,MINOS} has provided 
a compelling reason to extend the minimal version of the Standard Model~(SM).
Among possible ways to accommodate the neutrino mass, the introduction of  
gauge-singlet fermions (right-handed neutrinos) is one of the most 
natural and popular methods.
Similarly, the accumulated evidence for an unknown substance filling our Universe,
so-called dark matter~\cite{RotCurv,GravLens,WMAP,Planck}, has also provided an insight 
into physics beyond the SM.
A new particle species which is stable, neutral, and non-baryonic excellently fits
the observed nature of dark matter.
Interestingly, both of these two indicators may point to the existence of gauge-singlet 
fields beyond the SM. 
It is quite tempting to address these compelling issues by introducing 
the right-handed neutrinos and identifying one of them as the dark matter particle.

Along this line, the simplest extension of the SM is just
adding the right-handed neutrinos. 
In fact, three generations of right-handed neutrinos are able to 
accommodate the neutrino masses and mixing, dark matter, and baryon asymmetry of the 
Universe~\cite{nuMSM}.
Another simple example is the model in Ref.~\cite{Okada}, where a gauge-singlet scalar 
which breaks $U(1)_{B-L}$ gauge symmetry is introduced in addition to the three 
right-handed neutrinos.
Besides their simplicity, a virtue of these frameworks is that they can be free from the 
gauge hierarchy problem.
Since no intermediate scale between the electroweak/TeV and the Planck scale
is introduced in these models,
the problematic quadratic divergence in the Higgs mass can be a matter of the high-energy
boundary at which the theory becomes invalid~\cite{Bardeen,others}.
Since we do not know what sorts of boundary condition nature can realize,
we may postpone a fundamental solution of the gauge hierarchy problem and
attack compelling issues such as neutrino mass and dark matter within physics up to
the TeV scale. 
It is interesting to explore renormalizable theories beyond the SM
that stand on the same footing as the SM.

If this idea is further pushed with a few more new scalar fields,
one meets with a group of models, the so-called radiative seesaw with right-handed 
neutrinos~\cite{Krauss,Ma,Aoki}.
In this scenario, the usual neutrino Yukawa couplings are forbidden by some symmetries
so that the neutrino masses show up at some loop level.
Unlike the usual seesaw mechanism, the right-handed neutrinos do not mix with
the left-handed ones and they are separated from the electroweak interactions 
at the tree level.
The symmetries that forbid the Yukawa couplings make the lightest particle
running in the loop stable, providing a good candidate for dark matter. 

However, one may argue that there is an unsatisfying point which is common in 
the radiative models.
In many cases, a discrete symmetry such as $\mathbb{Z}_2$ is imposed to forbid 
the Yukawa couplings and stabilize the dark matter candidate.
The use of a global discrete symmetry seems rather ad-hoc in view of 
the great success of the gauge principle of the SM.
In addition, it has been argued that in the context of quantum gravity global 
symmetries are unnatural~\cite{Giddings}.
These facts motivate us to upgrade the global discrete symmetries
to local ones.
A simple realization of this idea, for example, is to derive $\mathbb{Z}_2$ 
symmetry as a remnant of a gauge symmetry broken at some high-energy 
scale~\cite{Krauss:1988zc}.
Along this line, various types of gauge symmetries such as $SU(2)'$~\cite{Walker},
$U(1)_{B-L}$~\cite{Nakayama,Ibe}, and $U(1)'$~\cite{Kubo:2006rm,Batell,Chang,Law,
Ma:2013yga} have been studied. 

In this work, we present yet another possibility for $U(1)'$.
We consider a $U(1)'$ gauge symmetry which acts on 
the three generations of the right-handed neutrinos $\nu_{R_i}\,(i=1,2,3)$
\footnote{We assume unknown physics governing the generation structure universally 
determines the number of the fermion families in the visible and hidden sectors.} as 
a symmetry behind the stability of dark matter.
The usual SM particles are neutral under this $U(1)'$.
For the theory being anomaly free, the charges for $\nu_{R_i}$ are uniquely determined
up to an overall factor and labeling of the generation;
\begin{center}
\renewcommand{\arraystretch}{1.25}
\begin{tabular}{cccc}\hline\hline
& $\nu_{R_1}$ & $\nu_{R_2}$ & $\nu_{R_3}$ \\\hline
$U(1)'$ & $0$ & $q$ & $-q$ \\\hline
\end{tabular}
\end{center}
If the charges are rational, there is no ``chiral solution'', in which 
all the charges are finite and distinct from each other in their absolute values, 
i.e., $q_i\neq0$ and $q_i+q_j\neq0$ for any $i,j$~\cite{Batra:2005rh}
\footnote{The conditions for the anomaly cancellation are 
$q_1^3 + q_2^3 + q_3^3 =0$ (and $q_1 + q_2 + q_3 =0$).
If $q_i$ are rational, we may regard $q_i$ are integer since the overall normalization
does not matter.
Fermat's Last Theorem ($n =3$) tells us that there is no solution
for the condition.}.
It is shown that the case of the four right-handed neutrinos also has only vector-like 
solutions.
For the chiral solutions, at least five fermions are needed~\cite{Batra:2005rh}.

Under the charge assignment shown above, both tree and loop-level
contributions should be important in the neutrino mass (matrix) at low energy. 
Since $\nu_{R_1}$ is neutral under $U(1)'$, it has the Yukawa couplings with
the left-handed leptons and the Higgs doublets, so that 
the canonical seesaw mechanism in the first place produces a neutrino mass matrix 
of rank 1. 
The other right-handed neutrinos $\nu_{R_{2,3}}$ are responsible for the correction 
term needed to increase the rank of the whole neutrino mass matrix.
This correction term can show up in the same manner as radiative seesaw models.

As the scalar particles needed to have a viable loop correction, 
we consider a $SU(2)$ doublet scalar $\eta$ as a simple example in order to
obtain the correction at one loop.
The radiative seesaw model based on $\eta$ and the right-handed neutrinos has 
originally been proposed in Ref.~\cite{Ma}. 

As in many radiative seesaw models, the lightest of the neutral particles running 
in the loop becomes a good candidate for dark matter.
We assume that one of the right-handed neutrinos plays this role.
Besides $t$-channel exchange of $\eta$, the dark matter particle can communicate with 
the SM particles through $s$-channel exchange of neutral scalar particles. 
These scalars originate from the standard Higgs doublet $\Phi$ and a new scalar field 
$\phi$ that  breaks $U(1)'$ by its vacuum expectation value (VEV).
In the early Universe, the right-handed neutrinos are thermalized and then annihilate 
into the SM particles predominantly by this process.
The relic abundance does not rely on the $t$-channel exchange of $\eta$ in our model.
Hence the model is free from the tension between the correct relic abundance and 
lepton flavor violation (LFV) that the models of similar type to Ref.~\cite{Ma} 
often suffer~\cite{Suematsu:2009ww}.

The layout of this paper is as follows.
In Section~\ref{model} , we present in detail a concrete model,
where the mass and flavor structure of the induced neutrino mass matrix and 
their phenomenology will be the main focus.
In Section~\ref{darkmatter}, we discuss the physics of dark matter, 
including the relic abundance and a perspective on direct detection.
In Section~\ref{summary} we summarize our result and finish with a discussion.
\section{The model}
\label{model}
We introduce three right-handed neutrinos $\nu_{R_i}\, (i = 1,2,3)$ and
the $SU(2)$ doublet scalar $\eta$ which has the same weak charges as the
usual Higgs doublet $\Phi$. 
The new particles are charged under the new $U(1)'$ gauge symmetry which 
is broken at the TeV scale.
The particle contents and the quantum numbers under the weak interactions
and $U(1)'$ are summarized as follows.
\begin{center}
\renewcommand{\arraystretch}{1.25}
\begin{tabular}{ccccccc}\hline\hline
& $\nu_{R_1}$ & $\nu_{R_2}$ & $\nu_{R_3}$ &$\eta$ & $\phi$ & $\xi$ \\\hline
$SU(2)\times U(1)_Y$ & $(1,0)$ & $(1,0)$ & $(1,0)$ & $({\bf 2},1/2)$ & $(1,0)$ & $(1,0)$ 
\\
$U(1)'$ & $0$ & $q$ & $-q$ & $q$ & $2q$ & $q$ \\\hline
\end{tabular}
\end{center}
The scalar $\phi$ is introduced to break $U(1)'$ down to $\mathbb{Z}_2$\footnote{The 
gauge transformation $\phi(x) \to e^{i\alpha(x)2q}\phi(x)$ becomes
trivial for $\alpha(x) = 2\pi /2q$. This direction works as an odd parity for
the fields with charge $\pm q$.}.
Another scalar $\xi$ plays a role in the neutrino mass as we will see below.
The usual Standard Model fields are neutral under $U(1)'$.
The Lagrangian for the neutrino Yukawa sector is
\begin{eqnarray}
\mathcal{L} &\,=\,& y_{1 \alpha} \nu_{R_1}^\dag L_\alpha \Phi 
\,+\, y_{2 \alpha} \nu_{R_2}^\dag L_\alpha \eta 
\nonumber\\
&& +\,\frac{1}{2}M_1 \nu_{R_1}^{\rm T} \epsilon  \nu_{R_1} 
+ \frac{1}{2}(\nu_{R_2}^{\rm T},\nu_{R_3}^{\rm T})
\renewcommand{\arraystretch}{1.25}
\begin{pmatrix}
g_1 \phi^* & \widetilde{M}_2 \\
\widetilde{M}_2 & g_2 \phi \\
\end{pmatrix}
\begin{pmatrix}
\epsilon \, \nu_{R_2} \\
\epsilon \, \nu_{R_3} \\
\end{pmatrix}
\,+\, {\rm h.c.},
\label{Lag1}
\end{eqnarray}
where $L_\alpha \, (\alpha = e,\mu,\tau)$ are the left-handed lepton doublets,
$y_{1 \alpha}$, $y_{2 \alpha}$, $g_1$ and $g_2$ are dimensionless constants,
and $M_1$ and $\widetilde{M}_2$ are the Majorana  mass parameters.
We assume that $M_1$ and $\widetilde{M}_2$ take values around or less than TeV
to avoid fine tuning in the scalar masses.
Fermions are denoted by two-component spinors, and $\epsilon$ stands for 
the $2\times 2$ antisymmetric tensor on that spinor space. 

After $\phi$ getting a VEV $\langle \phi \rangle$, 
$\nu_{R_2},\nu_{R_3}$ are mixed to compose the mass eigenstates.
In terms of the mass eigenstates, the Yukawa couplings and the Majorana mass terms 
for the right-handed neutrinos
become\footnote{A similar Lagrangian is discussed in Ref.~\cite{Kubo:2006rm} 
where $U(1)'$ charges
are assigned not only to right-handed neutrinos but also to quarks.}
\begin{eqnarray}
\mathcal{L}'&\,=\,& y_{1 \alpha}\nu_{R_1}^\dag L_\alpha  \Phi
+ y_{2 \alpha}\left(\,\cos\theta N_2^\dag  + \sin\theta N_3^\dag \,\right) L_\alpha
\eta\nonumber\\
&&\,+\,\frac{1}{2}M_1 \nu_{R_1}^{\rm T} \epsilon  \nu_{R_1} 
\,+\,\frac{1}{2}M_2 N_2^{\rm T} \epsilon  N_2
\,+\,\frac{1}{2}M_3 N_3^{\rm T} \epsilon  N_3
\,+\, {\rm h.c.},
\label{Lag2}
\end{eqnarray}
where $N_{2,3}$ are related with the original fields as
\begin{eqnarray}
\begin{pmatrix}
\nu_{R_2} \\
\nu_{R_3} \\
\end{pmatrix}
\,=\, 
\begin{pmatrix}
\cos\theta & \sin\theta \\
-\sin\theta & \cos\theta \\
\end{pmatrix}
\begin{pmatrix}
N_2 \\
N_3 \\
\end{pmatrix}
\end{eqnarray}
with
\begin{eqnarray}
\tan2\theta = \frac{2\widetilde{M}_2}{(g_2 - g_1)\langle \phi \rangle},
\end{eqnarray}
\begin{eqnarray}
M_2 &=& \frac{1}{2}\left( \, (g_1 + g_2)\langle \phi \rangle - 
\sqrt{(g_1 - g_2)^2\langle \phi \rangle^2 + 4\widetilde{M_2}^2} \right),\nonumber\\
M_3 &=& \frac{1}{2}\left( \, (g_1 + g_2)\langle \phi \rangle + 
\sqrt{(g_1 - g_2)^2\langle \phi \rangle^2 + 4\widetilde{M_2}^2}\right).
\end{eqnarray}
Here we have assumed that $\langle \phi \rangle$, $g_{1,2}$ and $\widetilde{M_2}$
are real for simplicity.
From Eq.~(\ref{Lag2}), it is clear that the lightest of $N_2$, $N_3$ and $\eta$
becomes stable and can be a dark matter candidate due to the remnant $\mathbb{Z}_2$ 
symmetry. 

The scalar potential is given by
\begin{eqnarray}
V &\,=\,& 
+\,m_{\Phi}^2 \Phi^\dag \Phi \,+\, m_{\eta}^2 \eta^\dag \eta
\,+\, m_{\phi}^2 \phi^* \phi\,+\, m_{\xi}^2 \xi^* \xi
\nonumber\\
&& +\,(\, \mu_1 \,\eta^\dag \Phi \xi  \,+\, \mu_2 \,\phi\, \xi^* \xi^* + {\rm h.c.})
\nonumber\\
&& +\lambda_1 (\Phi^\dag \Phi)^2 + \lambda_2 (\eta^\dag \eta)^2 + 
\lambda_3 (\phi^* \phi)^2 + \lambda_4 (\xi^* \xi)^2 \nonumber\\
&&+\lambda_5 (\Phi^\dag \Phi)(\eta^\dag \eta)
+ \lambda_6 (\Phi^\dag \eta)(\eta^\dag \Phi) + 
\lambda_7 (\Phi^\dag \Phi)(\phi^* \phi)
+\lambda_8 (\Phi^\dag \Phi)(\xi^* \xi)\nonumber\\
&&
+\lambda_9 (\eta^\dag \eta)(\phi^*\phi) + \lambda_{10}(\eta^\dag \eta)(\xi^*\xi)
+ \lambda_{11}(\phi^*\phi)(\xi^*\xi)
+ \lambda_{12}(\eta^\dag \Phi \phi \xi^* + {\rm h.c.}),
\end{eqnarray}
where $m_{\Phi, \eta,\phi,\xi}^2$ are the parameters of mass dimension two,
$\mu_{1,2}$ are of the mass dimension one, $\lambda_i\,(i =1,\cdots 12)$ are
dimensionless couplings.
The coupling for $\eta^\dag \Phi \phi \xi^*$ can be made real with no loss of 
generality.
As mentioned above, we assume that only $\Phi$ and $\phi$ develop VEV, 
following
$\langle \Phi \rangle = \big(\begin{smallmatrix}0 \\ v\end{smallmatrix}\big)$
with $v = 174\,{\rm GeV}$. 
The VEV $\langle \phi \rangle$ is assumed to be around $1\,{\rm TeV}$. 

\begin{figure}[t]
    \begin{center}
\scalebox{0.6}{\includegraphics{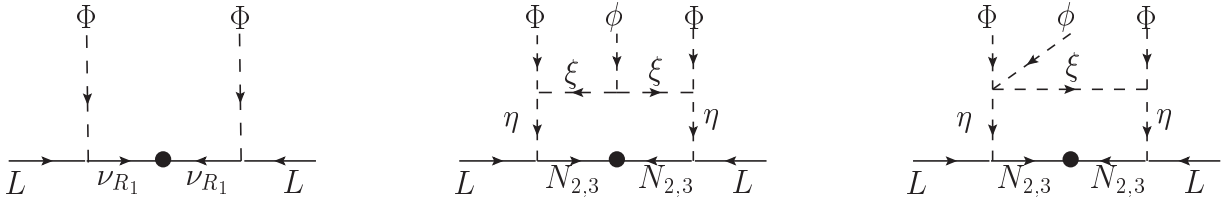}}
\end{center}
\caption{Feynman diagrams for the neutrino mass.}
\label{fig1}
\end{figure}

At low-energy scales, the neutrino mass receives 
two contributions;
\begin{eqnarray}
\mathcal{M}_\nu \,=\,\mathcal{M}_\nu^{\rm tree} + \mathcal{M}_\nu^{\rm rad},
\label{numass}
\end{eqnarray}
where $\mathcal{M}_\nu^{\rm tree}$ is from the Type I seesaw by $\nu_{R_1}$.
The other part $\mathcal{M}_\nu^{\rm rad}$ stems from the radiative corrections by
$N_2$ and $N_3$.
Fig.~\ref{fig1} shows the Feynman diagrams responsible for these mass terms.
Provided that $y_{1\alpha}v \ll M_1$, the tree-level contribution is given by
\begin{eqnarray}
(\mathcal{M}_\nu^{\rm tree})_{\alpha \beta} &\,\simeq\,& 
- y_{1 \alpha}y_{1 \beta}\frac{v^2}{M_1}
\label{a1}
\end{eqnarray}
in a good approximation.
The correction term turns out to be~\cite{Ma}
\begin{eqnarray}
(\mathcal{M}_\nu^{\rm rad})_{\alpha \beta} &\,\simeq\,& 
y_{2 \alpha}y_{2 \beta}\left( \frac{1}{4\pi^2} 
\frac{ v^2 \langle \phi \rangle}{M_*} \right)
\left(\frac{\cos^2\theta}{M_2}\,I\!\left[ \frac{M_2^2}{M_\eta^2} \right]
+ \frac{\sin^2\theta}{M_3}\,I\!\left[ \frac{M_3^2}{M_\eta^2} \right]
\right),
\label{a2}
\end{eqnarray}
where $I\!\left[ x \right] = \frac{x}{1-x}\left( 1 + \frac{x\ln x}{1-x} \right),$
and
\begin{eqnarray}
\frac{1}{M_*} \,= \, \frac{\mu_1^2 \mu_2 + \lambda_{12}\mu_1 m_\xi^2}{m_\xi^4},
\quad\quad M_\eta^2 \,=\,
m_\eta^2 + (\lambda_5 + \lambda_6)v^2 + \lambda_9 \langle \phi \rangle^2.
\label{mstar}
\end{eqnarray}
Here we have assumed that $M_{2,3} < M_\eta$ and $m_\xi$ is the largest 
of all the other mass parameters in the model.

The whole mass matrix in Eq.~(\ref{numass}) has rank 2, so that the neutrino mass
spectrum becomes either normal~(NH) or inverted hierarchical~(IH).
For IH, however, the two contributions $(\mathcal{M}_\nu^{\rm tree})$ and 
$(\mathcal{M}_\nu^{\rm rad})$ must be tuned at 1\% level.
Although this is possible, it might require an additional explanation for what
conspiracy is behind it.
In what follows, we stick to NH, postulating
$(\mathcal{M}_\nu^{\rm tree}) \approx m_3 = \sqrt{\Delta m^2_{31}}$ and
$(\mathcal{M}_\nu^{\rm rad}) \approx m_2 = \sqrt{\Delta m^2_{21}}$ as a natural 
possibility.

An illustrative example of fitting the neutrino masses and mixing is obtained by 
identifying the Yukawa couplings $y_{2\alpha}$ and $y_{1\alpha}$ as
\begin{eqnarray}
y_{2\alpha} \,=\, 
(U_{\alpha 2})\,y_2,\quad\quad
y_{1\alpha} \,=\, (U_{\alpha 3})\,y_1
\end{eqnarray}
and $m_2$ and $m_3$ as
\begin{eqnarray}
&&m_2 \,=\, y_2^2 \left( \frac{1}{4\pi^2} 
\frac{ v^2 \langle \phi \rangle}{M_*} \right)
\left(\frac{\cos^2\theta}{M_2}\,I\!\left[ \frac{M_2^2}{M_\eta^2} \right]
+ \frac{\sin^2\theta}{M_3}\,I\!\left[ \frac{M_3^2}{M_\eta^2} \right] \right),
\nonumber\\
&&m_3 \,=\, y_1^2 \frac{v^2}{M_1},
\label{masses}
\end{eqnarray}
where $U_{\alpha i}$ are the Pontecorvo-Maki-Nakagawa-Sakata
(PMNS) matrix~\cite{PDG}.
With this choice, the neutrino mass matrix Eq.~(\ref{numass}) is diagonalized by
the PMNS matrix with the eigenvalues given in Eq.~(\ref{masses}) up to the Majorana phase.
Note that we may regard the Lagrangian~(\ref{Lag1}) and~(\ref{Lag2})
in the basis where the charged-lepton mass matrix is diagonal without loss
of generality.
By taking $M_{2,3} = 100\,{\rm GeV}$, $M_1 = M_\eta = \langle \phi \rangle 
= 1 \,{\rm TeV}$ and
$M_* = 10\,{\rm TeV}$ for example, one finds 
$y_2 \approx 3 \times 10^{-5}$ and $y_1 \approx 1 \times 10^{-6}$ for typical values
of $|\Delta m^2_{31}|$ and $\Delta m^2_{21}$~\cite{globalfit}.
Since the Yukawa couplings are small, the LFV signals and the annihilation rate
of dark matter via $t$-channel exchange of $\eta$ are small
for the mass parameters around the TeV scale.
 

Finally, we comment on the role of $\xi$.
If $\xi$ were absent, the radiative correction would be induced at
the two-loop level.
In this case, $1/M_*$ in Eq.~(\ref{a2}) is given by 
\begin{eqnarray}
\frac{1}{M_*} \,\sim \,
\frac{1}{16\pi^2}\frac{\sum_{\alpha,\beta}y_{1 \alpha}^2 y_{2 \beta}^2}{M_1}
\end{eqnarray}
instead of the one in Eq.~(\ref{mstar}).
The radiative contribution would be much smaller than the tree-level
contribution if $\xi$ were absent.
The $\xi$ field is needed to keep the correction term $(\mathcal{M}_\nu^{\rm rad})$ 
sizable.

\section{Dark Matter}
\label{darkmatter}
In Section~\ref{model}, it was shown that the lightest of $N_2$, $N_3$, and $\eta$
becomes stable and can be a dark matter candidate.
We assume that one of the right-handed neutrinos serves as dark matter.
With no loss of generality, we can set $M_2 < M_3$ so that $N_2$ becomes the stable 
dark matter particle.

\subsection{Relic density}

\paragraph{Annihilation via Yukawa coupling} 
Let us start with the annihilation channel $N_2 N_2 \to LL$ via $t$-channel exchange 
of $\eta$. 
As shown in Section~\ref{model}, the Yukawa coupling $y_2$ responsible for 
$N_2 N_2 \to LL$ must be small provided that the mass parameters in
the model are supposed to be around TeV.
One may argue that $y_2$ can be increased by increasing the mass parameters beyond
TeV or by allowing some hierarchy between mass parameters.
However, larger values of $y_2$ are already in conflict with null observation of LFV 
in the charged leptons.

\begin{figure}[t]
    \begin{center}
\scalebox{0.9}{\includegraphics{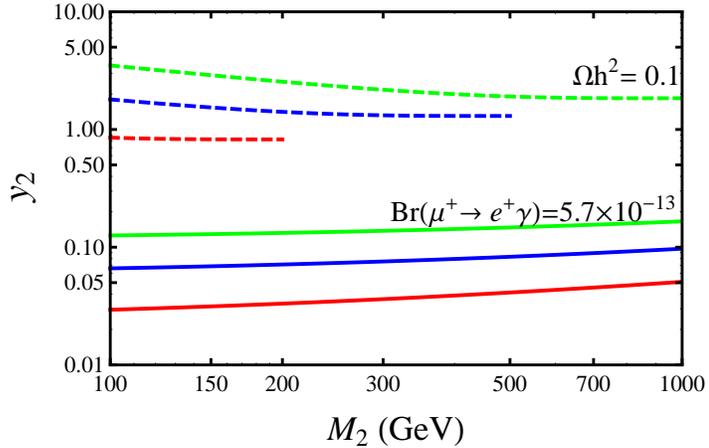}}
\end{center}
\caption{The contours in the $M_2$-$y_2$ plane for ${\rm Br}(\mu \to e \gamma)$ 
and the $N_2$ relic density $\Omega_N h^2$. 
The solid curves show the contours for 
${\rm Br}(\mu \to e \gamma) = 5.7\times 10^{-13}$, whereas
the dashed curves show the contours for $\Omega_N h^2 = 0.1$. 
The three curves in each constraint are the cases for 
$M_\eta = 200\,({\rm red}), 500\,({\rm blue}), 1000\,({\rm green}){\rm GeV}$ 
(from the bottom to top).}
\label{fig2}
\end{figure}

Fig.~\ref{fig2} shows the parameter space of $M_2$ and $y_2$ with
the contours for the branching ratio of $\mu^+ \to e^+ \gamma$ decay 
${\rm Br}(\mu\to e\gamma) = 5.7\times 10^{-13}$~\cite{MEG} and the $N_2$ 
relic density $\Omega_N h^2 = 0.1$~\cite{WMAP,Planck}. 
The plot shows the case where only the $t$-channel annihilation $N_2N_2 \to LL$ 
is available for the dark matter annihilation.
The solid curves show the contours for ${\rm Br}(\mu \to e \gamma)$, 
whereas the dashed curves show the contours for $\Omega_N h^2$. 
The three curves for each constraint are the cases of 
$M_\eta = 200\,({\rm red}), 500\,({\rm blue}), 1000\,({\rm green}){\rm GeV}$ 
(from the bottom to top).

The lepton mixing angles are assumed to be the best-fit values taken from 
Ref.~\cite{globalfit},
and $M_2 = M_3$ (or equivalently $\sin\theta =0$) is assumed for simplicity.
We do not present the explicite formulas for ${\rm Br}(\mu \to e \gamma)$
and $\Omega_N h^2$ (or the thermally averaged cross section for $N_2N_2 \to LL$) here,
but one can find the complete formulas in Ref.~\cite{Suematsu:2009ww}.
From Fig.~\ref{fig2}, it is clear that there is no parameter set which satisfies
both requirements from dark matter and LFV\footnote{We do not consider 
coannihilation $N_2N_3 \to LL$ here. 
The effect of this coannihilation in the model~\cite{Ma} is studied in 
Ref.~\cite{Suematsu:2009ww}.}.

\paragraph{Annihilation via Higgs scalars and $Z'$} 
Besides the $t$-channel exchange of $\eta$, 
the dark matter communicates with the thermal plasma through
the $s$-channel exchange of scalar particles (so-called Higgs portal~\cite{Patt:2006fw})
and the gauge boson $Z'$ whose mass is given by $M_{Z'} = 2\sqrt{2}g'q
\langle \phi \rangle$, where $g'$ is the gauge coupling of $U(1)'$.
We assume $\langle \phi \rangle \approx M_{Z'} \approx 1\,{\rm TeV}$ as a typical case.
We also assume that the kinetic mixing between $U(1)_Y$ and $U(1)'$ is small
such that it satisfies the experimental constraints~\cite{Jaeckel:2012yz}.
In this case the dominant contributions come from the Higgs portal since 
the larger the mediator's mass, the smaller the annihilation rate.
In addition, the $Z'$ boson weakly couples to the SM due to the small kinetic mixing.

There are two real scalar fields which serve as the Higgs portal.
As mentioned in Section~\ref{model}, the scalar doublet $\Phi$ and singlet $\phi$ 
develop VEV, and they are expanded as
\begin{eqnarray}
\Phi \,= \begin{pmatrix}
0 \\
\Phi_R + v
\end{pmatrix},\quad\quad
\phi = \phi_R + \langle \phi \rangle
\end{eqnarray}
in the unitary gauge.
These fields $\Phi_R$ and $\phi_R$ are related with the mass eigenstates $h$ and $H$
via
\begin{eqnarray}
\begin{pmatrix}
h \\
H \\
\end{pmatrix}\,=\,
\begin{pmatrix}
\cos\alpha & -\sin\alpha \\
\sin\alpha & \cos\alpha \\
\end{pmatrix}
\begin{pmatrix}
\Phi_R \\
\phi_R \\
\end{pmatrix},
\end{eqnarray}
where $\alpha$ is the mixing angle. 
We denote the masses of $h$ and $H$ as $M_h$ and $M_H$ ($M_h < M_H$), respectively. 
Through the $s$-channel exchange of $h$ and $H$, $N_2$ is thermalized in
the early Universe and then decoupled from the plasma while it is nonrelativistic.
If the masses of $\eta$, $\xi$ and $N_3$ are sufficiently larger than the dark matter
mass $M_2$, the annihilation rate is determined by the six parameters; 
$\alpha$, $M_h$, $M_H$, $v$, $\langle \phi \rangle$ and $\theta$.
This setup is similar to Ref.~\cite{Okada}
up to the mixing $\theta$ for the right-handed neutrinos.
\begin{figure}[t]
    \begin{center}
      \scalebox{0.9}{\includegraphics{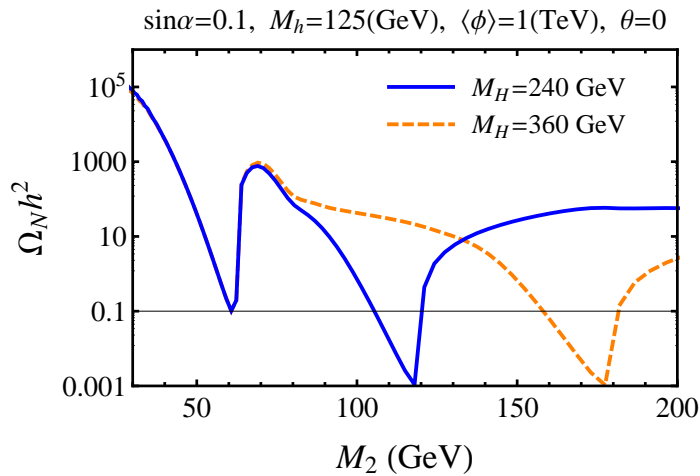}}
\end{center}
\caption{The thermal relic density of $N_2$ as a function of its mass $M_2$.}
\label{fig3}
\end{figure}

We implement the Lagrangian of our model into LanHEP~\cite{Semenov:2010qt}. 
To  calculate the relic density $\Omega_N h^2$ and the cross section for direct detection,
we use the micrOMEGAs software package~\cite{micrOMEGA}.
Fig.~\ref{fig3} shows the relic density $\Omega_N h^2$ as a function of
$M_2$ for $\sin\alpha = 0.1$, $M_h = 125\,{\rm GeV}$, $\langle \phi \rangle = 
1\,{\rm TeV}$, and $\sin\theta = 0$  for example.
The plot shows the two cases of the heavier Higgs mass; 
$M_H = 240\,{\rm GeV}$ and $360\,{\rm GeV}$.
It is seen that the observed abundance $\Omega_{\rm DM}h^2 = 0.1$ is 
achieved around the resonance $2M_2 \simeq M_{h,H}$.
The results are in good agreement with Ref.~\cite{Okada}.

In Fig.~\ref{fig3} we plot the relic density for $M_2 <200\,{\rm GeV}$
where the SM particles including $W^+W^-, ZZ$
are the dominant final states for the annihilation processes.
For $M_2 \gtrsim 200\,{\rm GeV}$,
the annihilation into the Higgs pairs is opened, so that 
the relic density decreases.
However, the relic density does not hit $\Omega_N h^2 = 0.1$ with this effect. 
The observed value $\Omega_N h^2 = 0.1$ can be obtained by the annihilation
into the non-SM particles such as $\eta$, $\xi$ and $N_3$.
For example, if these particles are lying around TeV, 
the correct relic density is achieved for $M_2 \gtrsim 700\,{\rm GeV}$ with
suitable choices of the parameters.

\begin{figure}[t]
    \begin{center}
\scalebox{0.65}{\includegraphics{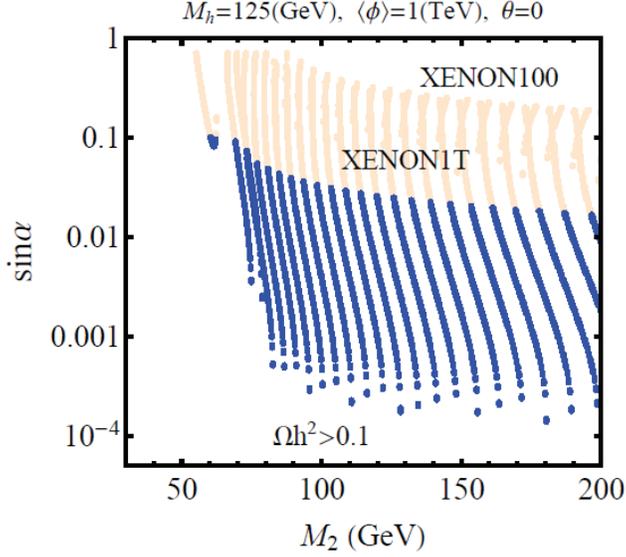}}
\end{center}
\caption{Scatter plot in $M_2$-$\sin\alpha$ space satisfying $\Omega_N h^2 = 0.1$ and
the direct detection bound from XENON100~\cite{XENON100}.
The direct detection and the relic density squeeze $\sin\alpha$ from larger and
smaller values, respectively.
For $\sin\alpha \lesssim 10^{-4}$, the relic density does not reach $\Omega_N h^2 = 0.1$
within $M_2 < 200\,{\rm GeV}$. The orange (light) points show the region 
corresponding to the cross sections XENON1T aims to probe in the future~\cite{XENON1T}.}
\label{fig4}
\end{figure}

\subsection{Direct detection}

The $t$-channel Higgs exchange induces dark matter scattering on nuclei. 
The elastic $N_2$-proton cross section $\sigma_p$
is given by~\cite{Ellis:2000ds}
\begin{eqnarray}
\sigma_p \,=\, 
\frac{4}{\pi}m_r^2 f_p^2,
\end{eqnarray}
where $m_r = m_p M_2/(m_p + M_2)$ is the reduced mass of the $N_2$-proton system
and
\begin{eqnarray}
f_p &\,=\,& \sum_{q = u,d,s} f_{Tq}^{(p)} \alpha_q \frac{m_p}{m_q}
+ \frac{2}{27}f_{TG}^{(p)} \sum_{q = c,b,t}  \alpha_q \frac{m_p}{m_q},
\nonumber\\
\alpha_q &\,=\,& \frac{1}{2}(g_1 \cos^2\theta + g_2 \sin^2\theta)y_q \sin 2\alpha
\left( \frac{1}{M_h^2} - \frac{1}{M_H^2} \right).
\end{eqnarray}
Here $f_{Tq}^{(p)}$ and $ f_{TG}^{(p)}$ are constants related to the proton
matrix elements, and $m_q$ and $y_q$ are the quark masses and the Yukawa couplings, 
respectively.
The cross section depends on the dark matter mass $M_2$ mainly through the coupling 
constants $g_{1,2}$ with
fixed values of $\langle \phi \rangle$ and $\widetilde{M}_2$.

Fig.~\ref{fig4} shows a scatter plot in the $M_2$-$\sin\alpha$ plane satisfying 
$\Omega_N h^2 = 0.1$ and the direct detection bound from XENON100~\cite{XENON100}.
In the plot we varied $M_H$ from $150\,{\rm GeV}$ to $500\,{\rm GeV}$ while
fixing the other parameters as $M_h = 125\,{\rm GeV}$, $\langle \phi \rangle = 
1\,{\rm TeV}$ and $\sin\theta = 0$ as a benchmark example.
The $N_2$-proton cross section is calculated in micrOMEGAs~\cite{micrOMEGA}.
Provided that the resonance conditions $2M_2 \simeq M_{h,H}$ hold, a wide parameter
space is still viable for the dark matter physics.
We also show the points (orange, light) to be probed by the future experiment 
XENON1T~\cite{XENON1T}.
It is seen that the mixing angle $\sin\alpha > 0.1$ will be sorted in the near future.

\section{Summary and discussions}
\label{summary}
Neutrino masses and dark matter have provided compelling reasons to extend the Standard
Model of elementary particle physics.
Assuming that three generations of right-handed neutrinos are responsible for 
these two deficiencies, we have discussed a $U(1)'$ gauge symmetry acting on them.
The $U(1)'$ breaks down at TeV and its remnant symmetry ensures the stability of 
one of the right-handed neutrinos $N_2$ that becomes a dark matter candidate.
For the anomaly being canceled, one right-handed neutrino must be neutral and the other 
two must have opposite charges under the $U(1)'$. 
This provides two contributions to the neutrino mass, i.e., a tree-level seesaw 
and a radiative correction with new scalar fields charged under the $U(1)'$ in the loops.
The neutrino masses and mixing are provided by these two contributions with sufficient
degrees of freedom to fit the observed data in neutrino oscillations.

In the early Universe, the dark matter candidate $N_2$ communicates with the thermal 
plasma via $t$ and $s$-channel exchange of the scalar and vector particles in the model.
Among them, the $s$-channel exchange of the scalar particles, the so-called Higgs portal, serves as the dominant annihilation channel.
By identifying the lightest scalar mediator as the SM like Higgs boson with a mass of $125\,{\rm GeV}$, we have searched viable parameter sets for dark matter phenomenology.
Making use of the resonances of the two scalar mediators, dark matter heavier than
$60\,{\rm GeV}$ can aquire annihilation cross sections sufficiently large for the correct 
relic density, without contradicting the recent direct detection experiments.

\vspace{3mm}

The LHC and other collider experiments may provide a glimpse into the extended Higgs 
sector. 
In our setup, the Higgs couplings to the SM particles are suppressed 
by $\cos\alpha$ compared to the SM case.
The recent LHC Higgs data combined with other experiments favor 
the range $[0.98,1.08]$ for 
the Higgs coupling to $W$ and $Z$ bosons relative to the SM~\cite{Falkowski:2013dza}.
This suggests $\sin\alpha 
\gtrsim 0.2$ is already disfavored.
Besides the overall suppression of the coupling, invisible decay may play a role.
If the dark matter mass is lighter than half of the Higgs mass, 
the Higgs particle decays into the dark matter.
Ref.~\cite{Kanemura:2011vm} studies a similar setup to our model and shows
that the branching ratio for the invisible decay reaches a few percent
with the maximal $\sin\alpha$.
A global fit to current data from ATLAS, CMS and Tevatron allows the invisible 
branching ratio up to $24$\% for a similar setup to the current one
\cite{Belanger:2013xza}.
The International Linear Collider may probe it down to a few percent
\cite{Schumacher:2003ss}.

The $Z'$ boson associated with the $U(1)'$ gauge symmetry may also leave interesting 
signals in experiments via the kinetic mixing with the $U(1)_Y$ gauge field.
Since the inert doublet $\eta$ is charged under both $U(1)_Y$ and $U(1)'$,  
the kinetic mixing is radiatively induced and the coefficient of the mixed 
kinetic term $\kappa$ can be around $10^{-3}$ even if it is vanishing at the tree level.
The kinetic mixing $\kappa \gtrsim 0.03$ for $10\,{\rm GeV}\lesssim M_{Z'} \lesssim 
1\,{\rm TeV}$ is disfavored by the data of $pp \to Z' \to \mu^+ \mu^-/e^+ e^-$ 
from ATLAS and CMS, and by the electroweak precision  constraints arising predominantly 
from $Z$ boson properties.
For $1\,{\rm GeV}\lesssim M_{Z'} \lesssim 10\,{\rm GeV}$, the measurements of
the invariant mass distribution of lepton pair in $e^+ e^- \to \gamma l^+ l^-$ with 
$\Upsilon$ resonances disfavor $\kappa \gtrsim 0.003$~(see, for instance, 
Ref.~\cite{Jaeckel:2012yz} and references therein).

\vspace{3mm}

Throughout the analysis in Section~\ref{darkmatter}, 
we have neglected the mixing $\theta$ among
two right-handed neutrinos charged under $U(1)'$.
Within this simple case, the $N_2N_2\phi$ coupling is real and the dark matter 
annihilates in a parity-conserving manner.
If $\theta$ is switched on, however, the $N_2N_2\phi$ coupling turns out to be complex,
and the dark matter can annihilate in a parity-violating way.
With this pseudo-Dirac coupling, the cross section for the direct detection can get
velocity suppression, so that evasion of the direct detection becomes much easier than
the case for the scalar coupling only~\cite{LopezHonorez:2012kv}.
The viable parameter space for the dark matter phenomenology is actually much
wider than the one presented in Section~\ref{darkmatter}.

Furthermore, if $\theta \neq 0$, the $U(1)'$ breaking VEV $\langle \phi \rangle$ can be 
small irrespective of the dark matter mass since $\theta \neq 0$ means the bare 
Majorana mass $\widetilde{M}_2 \neq 0$.
In the case where $\langle \phi \rangle \ll \widetilde{M}_2$, 
$\langle \phi \rangle$ can be smaller than TeV while keeping the dark matter mass
of $\mathcal{O}(100)\,{\rm GeV}$.
In this case the two right-handed neutrinos $N_2$ and $N_3$ compose a pseudo-Dirac
pair and a small mass splitting may be interesting in view of coannihilation and
baryogenesis.
Furthermore, if the VEV $\langle \phi \rangle$ is smaller than TeV, the Yukawa 
coupling $y_2$ gets larger than the benchmark value $\mathcal{O}(10^{-5})$ shown in 
Section~\ref{model}.
In this case, LFV such as $\mu^+ \to e^+ \gamma$ might be in reach of 
future experiments.

\subsection*{Acknowledgments}
D.S. acknowledges support by the International Max Planck Research School for Precision 
Tests of Fundamental Symmetries.
A.W. thanks the Young Researcher Overseas Visits Program for Vitalizing 
Brain Circulation Japanese in JSPS (No.~R2209).
\bigskip


\end{document}